\documentclass[pre,aps,preprint,showpacs,preprintnumbers,amsmath,amssymb,secnumroman,eqsecnum]{revtex4}

\usepackage{graphicx}
\usepackage{dcolumn}
\usepackage{bm}

\newcommand{\beq}{\begin{equation}}
\newcommand{\eeq}{\end{equation}}
\newcommand{\beqa}{\begin{eqnarray}}
\newcommand{\eeqa}{\end{eqnarray}}
\newcommand{\vc}[1]{\mbox{\boldmath $#1$}}

\newcommand{\vol}[1]{{\bf #1}}

\newcommand{\du}[1]{{\bf\sf #1}}

\begin{document}


\title{Swimming at small Reynolds number of a planar assembly of spheres in an incompressible viscous fluid with inertia}

\author{B. U. Felderhof}

 \email{ufelder@physik.rwth-aachen.de}
\affiliation{Institut f\"ur Theorie der Statistischen Physik \\ RWTH Aachen University\\
Templergraben 55\\52056 Aachen\\ Germany\\
}%

\date{\today}

\begin{abstract}
Translational and rotational swimming at small Reynolds number of a planar assembly of identical spheres immersed in an incompressible viscous fluid is studied on the basis of a set of equations of motion for the individual spheres. The motion of the spheres is caused by actuating forces and forces derived from a direct interaction potential, as well as hydrodynamic forces exerted by the fluid  as frictional and added mass hydrodynamic interactions. The translational and rotational swimming velocities of the assembly are deduced from momentum and angular momentum balance equations. The mean power required during a period is calculated from an instantaneous power equation. Expressions are derived for the mean swimming velocities and the mean power, valid to second order in the amplitude of displacements from the relative equilibrium positions. Hence these quantities can be evaluated for prescribed periodic displacements. Explicit calculations are performed for three spheres interacting such that they form an equilateral triangle in the rest configuration.

\end{abstract}

\pacs{47.15.G-, 47.63.mf, 47.63.Gd, 47.63.M-}
\maketitle
\section{\label{I}Introduction}

The mechanism of swimming of a body immersed in an infinite viscous fluid is not yet fully understood. Much progress has been made in the study of the Stokes limit \cite{1},\cite{2}, pertaining to the swimming of microorganisms, with neglect of inertia of the body and the fluid. The effect of fluid inertia can be characterized by a dimensionless viscosity \cite{3}, or equivalently by a scale number \cite{4}, given by the ratio of a typical body dimension to the range of viscous dissipation during a period. The Stokes limit corresponds to infinite dimensionless viscosity and vanishing scale number. In the following we study a model swimmer in the full range of scale number. The Reynolds number is assumed to be small, so that vortex shedding is neglected and the fluid flow is laminar. In the swimming of fish and humans and in the flying of birds the Reynolds number is large and vortex shedding is relevant \cite{5},\cite{6},\cite{6A}.

One can gain much insight by modeling the body as an assembly of spheres interacting directly, for example via harmonic springs, and with the effect of fluid motion embodied in frictional and added mass hydrodynamic interactions. The added mass of human bodies has been measured experimentally \cite{7}. In swimming the body is set in motion by actuating forces which vary periodically in time.

So far most model calculations in the Stokes limit have been made for linear chain structures with longitudinal motions along the chain. In particular, the linear three-sphere chain has been studied in detail \cite{8},\cite{9}. In earlier work we studied the effect of body and fluid inertia on the swimming of a linear three-sphere chain \cite{10},\cite{10A}. In the following we consider the more complicated situation of planar assemblies with motions restricted to a plane. The particular case of a triangular assembly in the Stokes limit was studied by Vladimirov \cite{11}.

In Sec. II the dynamics of the resistive-reactive model is formulated on the basis of Hamiltonian equations of motion with added frictional forces. The frictional hydrodynamic interactions can in principle be calculated in the Stokes limit \cite{12}. In practice one uses an Oseen \cite{13} or Rotne-Prager approximation \cite{14}. The added mass hydrodynamic interactions can in principle be found from potential theory \cite{15},\cite{16}. In practice one uses a dipole approximation \cite{10}. The spheres are also subject to direct forces, as specified in an interaction potential, and to actuating forces, driving the motion of the assembly. The total actuating force and torque are assumed to vanish.

The translational and rotational swimming velocity are found from momentum and angular momentum balance equations for the assembly. In Sec. III these equations are derived from the equations of motion for the individual spheres. In addition we derive an instantaneous power equation \cite{17}.

For simplicity we restrict attention to small amplitude motion, which implies small Reynolds number. The latter is a combination of the amplitude and the scale number. In Secs. IV and V we derive expressions for the mean translational and rotational swimming velocities, valid to second order in the amplitude of displacements from the equilibrium positions. In Secs. VI and VII we apply the theory to an assembly of three spheres, with rest configuration in the form of an equilateral triangle. In Sec. VIII we calculate the power, averaged over a period, to second order in the amplitude of displacements and optimize the mean translational swimming velocity for given power with respect to the stroke. For short mean distances between the spheres the optimal efficiency of swimming of the triangular configuration is substantially larger than that of a comparable three-sphere linear chain.

\section{\label{II}Dynamics of spheres centered in a plane}

We consider a system of $N$ identical spheres of radius $a$ and mass density $\rho_p$ immersed in a viscous incompressible fluid of shear viscosity $\eta$ and mass density $\rho$. The fluid is of infinite extent in all directions. We assume that at all times the centers of the spheres are located in the $xy$ plane of a Cartesian system of coordinates, so that the centers constitute a horizontal planar configuration. The dynamics of the system is governed by an interaction potential $V_{int}$, depending on the instantaneous configuration of centers, by actuating forces $(\vc{E}_1(t),...,\vc{E}_N(t))$, directed in the $xy$ plane and summing to zero total force and zero total torque at any time $t$, and by hydrodynamic forces exerted by the fluid. We assume that the spheres are freely rotating, so that the torque on any sphere vanishes. Further we assume that the hydrodynamic interactions can be approximated by Stokes friction, calculated from the Stokes equations, and by added mass effects, calculated from potential theory. We summarize the positions of centers in the $2N$-dimensional vector $\du{R}=(\vc{R}_1,...,\vc{R}_N)$, and the sphere momenta in the $2N$-dimensional vector $\du{p}=(\vc{p}_1,...,\vc{p}_N)$. The momenta are related to the velocities $\du{U}=(\vc{U}_1,...,\vc{U}_N)$ by
\begin{equation}
\label{2.1}\du{p}=\du{m}\cdot\du{U},
\end{equation}
where $\du{m}$ is the mass matrix, which depends on the relative positions of the spheres, so that it is invariant under translations of the whole assembly. Its tensor character determines the transformation under rotations of the assembly. The dynamics of the system is assumed to be governed by the approximate equations of motion \cite{10}
\begin{equation}
\label{2.2}\frac{d\du{R}}{dt}=\du{U},\qquad\frac{d\du{p}}{dt}=-\frac{\partial\mathcal{K}}{\partial\du{R}}-\vc{\zeta}\cdot\du{U}-\frac{\partial V_{int}}{\partial\du{R}}+\du{E},
\end{equation}
where the kinetic energy $\mathcal{K}$ is given by
\begin{equation}
\label{2.3}\mathcal{K}=\frac{1}{2}\;\du{p}\cdot\du{w}\cdot\du{p},
\end{equation}
with inverse mass matrix $\du{w}=\du{m}^{-1}$. The derivative with respect to positions in Eq. (2.2) is to be taken at constant momenta $\du{p}$. The friction matrix $\vc{\zeta}$ and the interaction potential $V_{int}$ are invariant under translations of the assembly. The interaction potential is invariant under rotations and the friction matrix transforms according to its tensor character. We abbreviated $\du{E}=(\vc{E}_1,...,\vc{E}_N)$. In spite of the fact that the total actuating force and torque vanish, the assembly can experience a net translation and rotation which is identified as translational and rotational swimming.

In the absence of actuating forces the system comes to rest due to friction with the fluid. The rest situation corresponds to a solution of Eq. (2.2) with constant configuration $\du{R}_0$, which is a minimum of the potential energy $V_{int}$. In the rest configuration the center and relative positions are
\begin{equation}
\label{2.4}\vc{C}_0=\frac{1}{N}\sum^N_{j=1}\vc{R}_{0j},\qquad\vc{c}_{0j}=\vc{R}_{0j}-\vc{C}_0,\qquad j=1,...,N.
\end{equation}
In shorthand notation $\du{c}_0=\du{R}_0-\du{C}_0$. From the definitions it follows that
\begin{equation}
\label{2.5}\du{u}_\alpha\cdot\du{c}_0=0,\qquad (\alpha=x,y),
\end{equation}
where $\du{u}_x=(1,0,1,0,...,1,0)$ and $\du{u}_y=(0,1,0,1,...,0,1)$. With respect to fixed axes with unit vectors $\vc{e}_x,\vc{e}_y$
\begin{equation}
\label{2.6}\vc{c}_{0j}=b_{jx}\vc{e}_x+b_{jy}\vc{e}_y,\qquad j=1,...,N,
\end{equation}
where the constants $b_{jx},b_{jy}$ are determined by the potential energy. We call $\du{c}_0$ the basic configuration. By isotropy a configuration rotated about the center by angle $\varphi$ with vectors
\begin{equation}
\label{2.7}\vc{c}_{j}=b_{jx}\vc{e}_{\rho}+b_{jy}\vc{e}_{\varphi},\qquad j=1,...,N,
\end{equation}
with rotated axes
\begin{eqnarray}
\label{2.8}\vc{e}_{\rho}&=&\cos\varphi\;\vc{e}_x+\sin\varphi\;\vc{e}_y,\nonumber\\
\vc{e}_{\varphi}&=&-\sin\varphi\;\vc{e}_x+\cos\varphi\;\vc{e}_y
\end{eqnarray}
is also an equilibrium configuration.

The positions of the centers at time $t$ may be decomposed as
\begin{equation}
\label{2.9}\vc{R}_j(t)=\vc{C}_0+\int^t_0\vc{U}(t')\;dt'+\vc{c}_j(t)+\vc{d}_j(t),\qquad j=1,...,N,
\end{equation}
where $\vc{U}(t)$ is the center velocity at time $t$, and $\vc{d}_j(t)$ is the additional displacement from the equilibrium structure.
The instantaneous positions $\du{R}(t)$ have center $\vc{C}(t)$, given by the first two terms in Eq. (2.9). In shorthand notation
\begin{equation}
\label{2.10}\du{R}=\du{C}+\du{c}+\du{d}.
\end{equation}
From the definitions it follows that
\begin{equation}
\label{2.11}\du{u}_\alpha\cdot\du{d}(t)=0,\qquad (\alpha=x,y).
\end{equation}
The orientation $\varphi(t)$ of the equilibrium structure $\du{c}(t)$ may be defined conveniently from the instantaneous positions $\du{R}(t)$. We choose to define it such that the squared distance $\du{d}(t)\cdot\du{d}(t)$ is minimal for the orientation $\varphi(t)$.

The time-derivative of Eq. (2.7) yields
\begin{equation}
\label{2.12}\frac{d\vc{c}_{j}}{dt}=\Omega\;\vc{e}_z\times\vc{c}_j,\qquad j=1,...,N,
\end{equation}
with angular velocity
\begin{equation}
\label{2.13}\Omega=\frac{d\varphi}{dt}.
\end{equation}
In matrix notation this can be expressed as
\begin{equation}
\label{2.14}\frac{d\du{c}}{dt}=-\Omega\;\du{X}\cdot\du{c},
\end{equation}
where for $N=2$ the matrix $\du{X}$ reads
\begin{eqnarray}
\label{2.15}\du{X}=\left(\begin{array}{cccc}
0&1&0&0
\\-1&0&0&0\\0&0&0&1\\0&0&-1&0\end{array}\right).
\end{eqnarray}
The generalization to arbitrary $N$ is obvious. With this notation the time-derivative of Eq. (2.10) reads
\begin{equation}
\label{2.16}\du{U}=U_\beta\du{u}_\beta-\Omega\;\du{X}\cdot\du{c}+\dot{\du{d}}.
\end{equation}
Substituting this into Eq. (2.2) and requiring that the total actuating force and torque vanish we obtain equations of motion for $U_x,U_y,\Omega$, involving also the time-derivatives $\dot{\du{d}}$ and $\ddot{\du{d}}$ of the displacements $\du{d}(t)$.

\section{\label{III}Momentum, angular momentum, and power}

The requirement that the sum of actuating forces vanishes reads in abbreviated notation
\begin{equation}
\label{3.1}\du{u}_\alpha\cdot\du{E}=0,\qquad (\alpha=x,y).
\end{equation}
Similarly, the requirement that the total torque of actuating forces vanishes can be expressed as
\begin{equation}
\label{3.2}\du{R}\cdot\du{X}\cdot\du{E}=0.
\end{equation}
We can use these requirements to derive simple balance equations for the following components of total momentum and orbital angular momentum,
 \begin{equation}
\label{3.3}P_x=\du{u}_x\cdot\du{p},\qquad P_y=\du{u}_y\cdot\du{p},\qquad L_z=\du{R}\cdot\du{X}\cdot\du{p}.
\end{equation}
These quantities vary due to interaction with the fluid. From Eq. (2.2) we derive
 \begin{equation}
\label{3.4}\frac{dP_x}{dt}=-\du{u}_x\cdot\vc{\zeta}\cdot\du{U},\qquad\frac{dP_y}{dt}=-\du{u}_y\cdot\vc{\zeta}\cdot\du{U},
\qquad\frac{dL_z}{dt}=-\du{R}\cdot\du{X}\cdot\vc{\zeta}\cdot\du{U}.
\end{equation}
The kinetic energy and potential energy terms in Eq. (2.2) do not contribute to these expressions on account of translational invariance and isotropy.

The orbital angular momentum can be decomposed as
 \begin{eqnarray}
\label{3.5}L_z&=&L_z'+\du{C}\cdot\du{X}\cdot\du{P},\qquad\du{P}=\big(P_x\du{u}_x+P_y\du{u}_y\big)/N,\nonumber\\
L_z'&=&(\du{R}-\du{C})\cdot\du{X}\cdot\du{p},
\end{eqnarray}
where $L_z'$ is the angular momentum relative to the center of mass. This has the rate of change
 \begin{equation}
\label{3.6}\frac{dL_z'}{dt}=-(\du{R}-\du{C})\cdot\du{X}\cdot\vc{\zeta}\cdot\du{U}-\frac{d\du{C}}{dt}\cdot\du{X}\cdot\du{P}.
\end{equation}
The right-hand side depends only on velocities and relative positions.

By use of $\du{p}=\du{m}\cdot\du{U}$ in Eq. (3.4) and substitution of Eq. (2.16) we derive coupled equations for the velocity components $U_x,U_y,\Omega$ involving the
time-derivatives $\dot{\du{d}}$ and $\ddot{\du{d}}$. The actuating forces $\du{E}(t)$ do not occur explicitly in these equations and this allows a kinematic point of view in which the velocities $U_x,U_y,\Omega$ are determined from the equations for prescribed displacements $\du{d}(t)$.

We define the Hamiltonian $\mathcal{H}$ as
 \begin{equation}
\label{3.7}\mathcal{H}=\mathcal{K}+V_{int}.
\end{equation}
From Eq. (2.2) we find for its time-derivative
 \begin{equation}
\label{3.8}\frac{d\mathcal{H}}{dt}=-\mathcal{D}+\du{E}\cdot\du{U},
\end{equation}
with rate of dissipation
 \begin{equation}
\label{3.9}\mathcal{D}=\du{U}\cdot\vc{\zeta}\cdot\du{U}.
\end{equation}
This may be called the instantaneous power equation \cite{17}.

In periodic swimming the time-average of the rate of dissipation over a period equals the power used. We denote the average as
 \begin{equation}
\label{3.10}\overline{\mathcal{D}}=\frac{1}{\tau}\int^\tau_0\mathcal{D}(t)\;dt,
\end{equation}
where $\tau$ is the period. From Eq. (3.8) we see that the mean rate of dissipation equals the power, i.e. the work performed by the actuating forces during a period,
 \begin{equation}
\label{3.11}\overline{\mathcal{D}}=\overline{\du{E}\cdot\du{U}}.
\end{equation}
In the same way we see from Eqs. (3.4) and (3.6)
 \begin{eqnarray}
\label{3.12}\du{u}_x\cdot\overline{\vc{\zeta}\cdot\du{U}}=0,\qquad\du{u}_y\cdot\overline{\vc{\zeta}\cdot\du{U}}&=&0,\nonumber\\
\overline{(\du{R}-\du{C})\cdot\du{X}\cdot\vc{\zeta}\cdot\du{U}}+\overline{\frac{d\du{C}}{dt}\cdot\du{X}\cdot\du{P}}&=&0.
\end{eqnarray}
The first two equations show that in periodic swimming the mean drag vanishes. In the next section we investigate the third equation for small amplitude motion.

\section{\label{IV}Bilinear theory}
In the following we consider an assembly with small deviations from an equilibrium configuration $\du{c}_0$. The averaged Eqs. (3.12) are solved by formal expansion in powers of the displacements $\du{d}(t)$. We include terms up to second order. The displacements are assumed to vary harmonically in time at frequency $\omega=2\pi/\tau$.

To second order the first two equations (3.12) read
\begin{equation}
\label{4.1}\du{u}_\alpha\cdot\overline{\vc{\zeta}^{(1)}\cdot\du{U}^{(1)}}+\du{u}_\alpha\cdot\vc{\zeta}^0\cdot\overline{\du{U}^{(2)}}=0,
\qquad (\alpha=x,y),
\end{equation}
where $\vc{\zeta}^0$ is the friction matrix of the equilibrium configuration, which is time-independent. Similarly, to second order the third equation reads
\begin{equation}
\label{4.2}\du{c}_0\cdot\du{X}\cdot\overline{\vc{\zeta}^{(1)}\cdot\du{U}^{(1)}}+\overline{(\du{R}^{(1)}-\du{C}^{(1)})\cdot\du{X}\cdot\vc{\zeta}^0\cdot\du{U}^{(1)}}
+\du{c}_0\cdot\du{X}\cdot\vc{\zeta}^0\cdot\overline{\du{U}^{(2)}}+\overline{\frac{d\du{C}^{(1)}}{dt}\cdot\du{X}\cdot\du{P}^{(1)}}=0.
\end{equation}
In these equations we can put from Eq. (2.16)
\begin{equation}
\label{4.3}\overline{\du{U}^{(2)}}=\overline{U^{(2)}_\beta}\du{u}_\beta-\overline{\Omega^{(2)}}\du{X}\cdot\du{c}_0,
\end{equation}
where we have used $\overline{\Omega^{(1)}\du{X}\cdot\du{c}^{(1)}}=0$, as follows from
\begin{equation}
\label{4.4}\du{c}^{(1)}=-\varphi^{(1)}\du{X}\cdot\du{c}_0,\qquad\Omega^{(1)}=d\varphi^{(1)}/dt.
\end{equation}
This shows that the second order mean velocities $\overline{U^{(2)}_x},\overline{U^{(2)}_y},\overline{\Omega^{(2)}}$ can be calculated from time-averaged products of first order quantities. We use the abbreviation
\begin{equation}
\label{4.5}\hat{\du{c}}_0=a^{-1}\du{c}_0\cdot\du{X}=-a^{-1}\du{X}\cdot\du{c}_0.
\end{equation}
From Eq. (4.1) we find
\begin{equation}
\label{4.6}Z^0_{\alpha\beta}\overline{U^{(2)}_\beta}+Z^0_{\alpha c}a\overline{\Omega^{(2)}}=\overline{\mathcal{I}^{(2)}_{T\alpha}}\qquad (\alpha=x,y),
\end{equation}
with friction elements
\begin{equation}
\label{4.7}Z^0_{\alpha\beta}=\du{u}_\alpha\cdot\vc{\zeta}^0\cdot\du{u}_\beta,\qquad Z^0_{\alpha c}=\du{u}_\alpha\cdot\vc{\zeta}^0\cdot\hat{\du{c}}_0,
\end{equation}
and mean second order translational impetus
\begin{equation}
\label{4.8}\overline{\mathcal{I}^{(2)}_{T\alpha}}=-\du{u}_\alpha\cdot\overline{\vc{\zeta}^{(1)}\cdot\du{U}^{(1)}},\qquad (\alpha=x,y).
\end{equation}

We write the second term in Eq. (4.2) as
\begin{equation}
\label{4.9}\overline{(\du{R}^{(1)}-\du{C}^{(1)})\cdot\du{X}\cdot\vc{\zeta}^0\cdot\du{U}^{(1)}}=\overline{(\du{R}^{(1)}-\du{C}^{(1)})\cdot\du{X}\cdot{\vc{\zeta}^0}'\cdot\du{U}^{(1)}}
+\zeta_0\overline{(\du{R}^{(1)}-\du{C}^{(1)})\cdot\du{X}\cdot\du{U}^{(1)}},
\end{equation}
where
\begin{equation}
\label{4.10}{\vc{\zeta}^0}'=\vc{\zeta}^0-\zeta_0\du{I},\qquad\zeta_0=6\pi\eta a,
\end{equation}
with $2N\times 2N$ unit matrix $\du{I}$. We call the average in the last term in Eq. (4.9) the generalized ellipticity. It is a sum of ellipticities of the single particle displacements, but it also has pair contributions. We write the term in the form
\begin{equation}
\label{4.11}\zeta_0\overline{(\du{R}^{(1)}-\du{C}^{(1)})\cdot\du{X}\cdot\du{U}^{(1)}}=\zeta_0a^2\omega\;\overline{\mathcal{E}}=(1-\alpha)Z^0_{cc}a^2\overline{\Omega^{(2)}}
\end{equation}
with rotational friction coefficient
\begin{equation}
\label{4.12}Z^0_{cc}=\hat{\du{c}}_0\cdot\vc{\zeta}^0\cdot\hat{\du{c}}_0.
\end{equation}
The coefficient $\alpha$ and the second order rotational velocity $\overline{\Omega^{(2)}}$ are to be determined.
The remaining terms in Eq. (4.2) yield
\begin{equation}
\label{4.13}Z^0_{\alpha c}\overline{U^{(2)}_\alpha}+\alpha Z^0_{cc}a\overline{\Omega^{(2)}}=\overline{\mathcal{I}^{(2)}_{R}},
\end{equation}
with mean second order rotational impetus
\begin{equation}
\label{4.14}\overline{\mathcal{I}^{(2)}_{R}}=-\hat{\du{c}}_0\cdot\overline{\vc{\zeta}^{(1)}\cdot\du{U}^{(1)}}
-a^{-1}\overline{(\du{R}^{(1)}-\du{C}^{(1)})\cdot\du{X}\cdot{\vc{\zeta}^0}'\cdot\du{U}^{(1)}}-a^{-1}\overline{\frac{d\du{C}^{(1)}}{dt}\cdot\du{X}\cdot\du{P}^{(1)}}.
\end{equation}
In the first term in Eq. (4.13) we have used the symmetry of the friction matrix $\vc{\zeta}^0$. From Eqs. (4.6) and (4.13) we can solve for $\overline{U^{(2)}_x},\overline{U^{(2)}_y},a\overline{\Omega^{(2)}}$ in terms of $\overline{\mathcal{I}^{(2)}_{Tx}},\overline{\mathcal{I}^{(2)}_{Ty}},\overline{\mathcal{I}^{(2)}_{R}}$. Finally the coefficient $\alpha$ can be determined from Eq. (4.11).

\section{\label{V}Mean translational and rotational impetus}

In this section we study the mean translational and rotational impetus defined in Eqs. (4.8) and (4.14) in more detail.
The first order friction matrix can be expressed as
\begin{equation}
\label{5.1}\vc{\zeta}^{(1)}=a\varphi^{(1)}\hat{\du{c}}_0\cdot\nabla\vc{\zeta}\big|_0+\du{d}\cdot\nabla\vc{\zeta}\big|_0,
\end{equation}
where $\nabla$ is the gradient operator in $2N$-dimensional configuration space. From Eq. (2.16) we find for the first order velocity vector and the corresponding position vector
\begin{equation}
\label{5.2}\du{R}^{(1)}=C_\beta^{(1)}\du{u}_\beta+a\varphi^{(1)}\hat{\du{c}}_0+\du{d},\qquad\du{U}^{(1)}=U_\beta^{(1)}\du{u}_\beta+a\Omega^{(1)}\hat{\du{c}}_0+\dot{\du{d}}.
\end{equation}
Here we use first order equations derived from Eqs. (3.4) and (3.6). From Eq. (3.4) we find
\begin{equation}
\label{5.3}\mathcal{M}^0_{\alpha\beta}\frac{dU^{(1)}_\beta}{dt}+\mathcal{M}^0_{\alpha c}a\frac{d\Omega^{(1)}}{dt}+\du{u}_\alpha\cdot\du{m}^0\cdot\ddot{\du{d}}=-Z^0_{\alpha\beta}U^{(1)}_\beta-Z^0_{\alpha c}a\Omega^{(1)}-\du{u}_\alpha\cdot\vc{\zeta}^0\cdot\dot{\du{d}},
\end{equation}
with mass elements
\begin{equation}
\label{5.4}\mathcal{M}^0_{\alpha\beta}=\du{u}_\alpha\cdot\du{m}^0\cdot\du{u}_\beta,\qquad \mathcal{M}^0_{\alpha c}=\du{u}_\alpha\cdot\du{m}^0\cdot\hat{\du{c}}_0.
\end{equation}
From Eq. (3.6) we find
\begin{equation}
\label{5.5}\mathcal{M}^0_{c\beta}\frac{dU^{(1)}_\beta}{dt}+\mathcal{M}^0_{cc}a\frac{d\Omega^{(1)}}{dt}
+\hat{\du{c}}_0\cdot\du{m}^0\cdot\ddot{\du{d}}=-Z^0_{c\beta}U^{(1)}_\beta-Z^0_{cc}a\Omega^{(1)}-\hat{\du{c}}_0\cdot\vc{\zeta}^0\cdot\dot{\du{d}},
\end{equation}
with mass elements
\begin{equation}
\label{5.6}\mathcal{M}^0_{c\beta}=\hat{\du{c}}_0\cdot\du{m}^0\cdot\du{u}_\beta= \mathcal{M}^0_{\beta c},\qquad \mathcal{M}^0_{cc}=\hat{\du{c}}_0\cdot\du{m}^0\cdot\hat{\du{c}}_0.
\end{equation}
These equations may be regarded as linear response equations determining the first order velocities $U^{(1)}_x,U^{(1)}_y,\Omega^{(1)}$ in terms of the displacements $\du{d}$. The equations can be solved by Fourier analysis. The equations for the complex Fourier coefficients read
\begin{eqnarray}
\label{5.7}\big[-i\omega\mathcal{M}^0_{\alpha\beta}+Z^0_{\alpha\beta}\big]U^{(1)}_{\beta\omega}+\big[-i\omega\mathcal{M}^0_{\alpha c}+Z^0_{\alpha c}\big] a\Omega^{(1)}_\omega&=&\du{u}_\alpha\cdot[\omega^2\du{m}^0+i\omega\vc{\zeta}^0]\cdot\du{d}_\omega,\nonumber\\
\big[-i\omega\mathcal{M}^0_{c\beta}+Z^0_{c\beta}\big]U^{(1)}_{\beta\omega}+\big[-i\omega\mathcal{M}^0_{cc}+Z^0_{cc}\big] a\Omega^{(1)}_\omega&=&\hat{\du{c}}_0\cdot[\omega^2\du{m}^0+i\omega\vc{\zeta}^0]\cdot\du{d}_\omega.
\end{eqnarray}
We introduce the shorthand notation $(\hat{U}^{(1)}_x,\hat{U}^{(1)}_y,\hat{U}^{(1)}_c)=(U^{(1)}_x,U^{(1)}_y,a\Omega^{(1)})$ and corresponding impedance vectors $\du{f}_\rho(\omega)$ with
\begin{equation}
\label{5.8}\du{f}_{x,y}(\omega)=(-i\omega\du{m}^0+\vc{\zeta}^0)\cdot\du{u}_{x,y},\qquad\du{f}_c(\omega)=(-i\omega\du{m}^0+\vc{\zeta}^0)\cdot\hat{\du{c}}_0.
\end{equation}
Then the solution of Eq. (5.7) can be expressed as
\begin{equation}
\label{5.9}\hat{U}^{(1)}_{\rho\omega}=i\omega Y_{\rho\sigma}(\omega)\du{f}_\sigma(\omega)\cdot\du{d}_\omega,
\end{equation}
with $3\times 3$ admittance matrix $\vc{Y}(\omega)$, or alternatively as
\begin{equation}
\label{5.10}\hat{U}^{(1)}_{\rho\omega}=i\omega\vc{\Psi}_\rho(\omega)\cdot\du{d}_\omega,\qquad\vc{\Psi}_\rho(\omega)=Y_{\rho\sigma}(\omega)\du{f}_\sigma(\omega).
\end{equation}
The elements of the vector $\vc{\Psi}_\rho(\omega)$ are dimensionless.

In the calculation of the time-averaged components of the impetus in Eqs. (4.8) and (4.14) we encounter bilinear expressions. The averages are evaluated conveniently in complex notation. For example
 \begin{equation}
\label{5.11}\overline{\du{d}U^{(1)}_\beta}=\frac{1}{2}\;\mathrm{Re}\;i\omega\du{d}^*_\omega\vc{\Psi}_\beta(\omega)\cdot\du{d}_\omega.
\end{equation}
The leading contribution to the time-average of the translational impetus in Eq. (4.8) takes the form
\begin{equation}
\label{5.12}-\du{u}_\alpha\cdot\overline{(\du{d}\cdot\nabla\vc{\zeta})\cdot\dot{\du{d}}}=\frac{1}{2}\;\mathrm{Re}\;[i\omega\du{d}^*_\omega\cdot\du{D}^\alpha\big|_0\cdot\du{d}_\omega],
\end{equation}
with derivative friction matrix
\begin{equation}
\label{5.13}\du{D}^\alpha=\vc{\nabla}\du{f}_\alpha,\qquad\du{f}_\alpha=\vc{\zeta}\cdot\du{u}_\alpha=\du{u}_\alpha\cdot\vc{\zeta},
\end{equation}
as introduced earlier \cite{9}. We write the complete expression as
\begin{equation}
\label{5.14}\overline{\mathcal{I}^{(2)}_{T\alpha}}=\frac{1}{2}\;\mathrm{Re}\;[i\omega\du{d}^*_\omega\cdot\breve{\du{D}}^\alpha\big|_0\cdot\du{d}_\omega],
\end{equation}
where the $2N\times 2N$ matrix $\breve{\du{D}}^\alpha$ differs from $\du{D}^\alpha$ by corrections coming from the remaining terms in Eqs. (5.1) and (5.2). We have
\begin{equation}
\label{5.15}\overline{\varphi^{(1)}\Omega^{(1)}}=0,
\end{equation}
so that there are four correction terms. It is convenient to use tensor notation and to define the $2N\times 2N\times 2N$ $F$-array as
\begin{equation}
\label{5.16}F_{jkl}=\frac{\partial\zeta_{jl}}{\partial x_k},
\end{equation}
where $x_k$ denotes the components of $\du{R}$. We define the corresponding contractions
\begin{eqnarray}
\label{5.17}
G_{\alpha\beta k}&=&u_{\alpha j}F_{jkl}u_{\beta l},\qquad G_{\alpha c k}=u_{\alpha j}F_{jkl}\hat{c}_{0l},\nonumber\\
H_{\alpha l}&=&u_{\alpha j}\hat{c}_{0k}F_{jkl},\qquad H_{cl}=\hat{c}_{0j}\hat{c}_{0k}F_{jkl},
\end{eqnarray}
where we use Einstein's summation convention for latin indices. By comparison with Eq. (5.13)
\begin{equation}
\label{5.18}G_{\alpha\beta k}=D^\alpha_{kl}u_{\beta l},\qquad G_{\alpha c k}=D^\alpha_{kl}\hat{c}_{0l}.
\end{equation}
The mean translational impetus can then be expressed as
\begin{equation}
{\label{5.19}\overline{\mathcal{I}^{(2)}_{T\alpha}}=\bigg[-D^\alpha_{kl}\overline{d_k\dot{d}_l}-H_{\alpha l}u_{\beta l}\overline{a\varphi^{(1)}U_\beta^{(1)}}+(H_{\alpha  k}-G_{\alpha c k})\overline{a\Omega^{(1)}d_k}-G_{\alpha\beta k}\overline{d_kU^{(1)}_\beta}\bigg]\bigg|_0,}
\end{equation}
where it is indicated that finally the coefficients must be evaluated at $\du{R}_0$. The averages can be evaluated in complex notation. Besides Eqs. (5.11) and (5.12) we have
\begin{eqnarray}
\label{5.20}\overline{a\varphi^{(1)}U_\beta^{(1)}}&=&\frac{1}{2}\;\mathrm{Re}\;\big[-i\omega\du{d}^*_\omega\cdot\vc{\Psi}^*_c(\omega)\vc{\Psi}_\beta(\omega)\cdot\du{d}_\omega\big],\nonumber\\
\overline{a\Omega^{(1)}d_k}&=&\frac{1}{2}\;\mathrm{Re}\;\big[i\omega d^*_{k\omega}\Psi_{cl}(\omega)d_{l\omega}\big].
\end{eqnarray}
Substituting from Eq. (5.10) and using complex notation we obtain the matrix in Eq. (5.14) as
\begin{equation}
\label{5.21}\breve{\du{D}}^\alpha(\omega)=\du{D}^\alpha+\du{H}_\alpha\cdot\du{u}_\beta\vc{\Psi}^*_c(\omega)\vc{\Psi}_\beta(\omega)-\du{G}_{\alpha\beta}\vc{\Psi}_\beta(\omega)
+(\du{H}_\alpha-\du{G}_{\alpha c})\vc{\Psi}_c(\omega).
\end{equation}

Next we consider the time-average of the rotational impetus, defined in Eq. (4.14). In generalization of Eq. (5.13) we need the rotational derivative friction matrix given by
\begin{equation}
\label{5.22}\du{D}^c=\vc{\nabla}\du{f}_c,\qquad\du{f}_c=\vc{\zeta}\cdot\hat{\du{c}}_0=\hat{\du{c}}_0\cdot\vc{\zeta}.
\end{equation}
In analogy with Eq. (5.14) we have
\begin{equation}
\label{5.23}\overline{\mathcal{I}^{(2)}_R}=\frac{1}{2}\;\mathrm{Re}\;[i\omega\du{d}^*_\omega\cdot\breve{\du{D}}^c(\omega)\big|_0\cdot\du{d}_\omega],
\end{equation}
with a matrix $\breve{\du{D}}^c(\omega)$. From the first term in Eq. (4.14) we obtain correction terms as in Eq. (5.21) with the subscript $\alpha$ replaced by $c$. The corresponding $G$-coefficients are
\begin{equation}
\label{5.24}
G_{c\beta k}=\hat{c}_{0j}F_{jkl}u_{\beta l}=G_{\beta ck},\qquad G_{cc k}=\hat{c}_{0j}F_{jkl}\hat{c}_{0l}.
\end{equation}
The time-average of the second term in Eq. (4.14) is
\begin{equation}
\label{5.25}
\overline{(\du{R}^{(1)}-\du{C}^{(1)})\cdot\du{X}\cdot{\vc{\zeta}^0}'\cdot\du{U}^{(1)}}=\overline{a\varphi^{(1)}\hat{\du{c}}_{0}\cdot\du{X}\cdot{\vc{\zeta}^0}'\cdot\du{U}^{(1)}}
+\overline{\du{d}\cdot\du{X}\cdot{\vc{\zeta}^0}'\cdot\du{U}^{(1)}}.
\end{equation}
The first term on the right is evaluated to
\begin{equation}
\label{5.26}
\overline{a\varphi^{(1)}\hat{\du{c}}_{0}\cdot\du{X}\cdot{\vc{\zeta}^0}'\cdot\du{U}^{(1)}}=-a^{-1}\du{c}_0\cdot{\vc{\zeta}^0}'\cdot\du{u}_\beta\overline{a\varphi^{(1)}U_\beta^{(1)}}
+a^{-1}\du{c}_0\cdot{\vc{\zeta}^0}'\cdot\overline{a\Omega^{(1)}\du{d}},
\end{equation}
with use of $\du{X}^2=-\du{I}$ and Eq. (5.15). The two averages are given by Eq. (5.20). The second term on the right in Eq. (5.25) is given by
\begin{equation}
\label{5.27}
\overline{\du{d}\cdot\du{X}\cdot{\vc{\zeta}^0}'\cdot\du{U}^{(1)}}=\frac{1}{2}\;\mathrm{Re}\;\big[i\omega d^*_{k\omega}X_{km}\big({f^0}'_{\beta m}\Psi_{\beta l}(\omega)+{f^0}'_{cm}\Psi_{c l}(\omega)-{\zeta^0}'_{ml}\big)d_{l\omega}\big].
\end{equation}
The time-average of the third term on the right in Eq. (4.14) vanishes on account of
\begin{equation}
\label{5.28}\overline{\frac{d\du{C}^{(1)}}{dt}\cdot\du{X}\cdot\du{P}^{(1)}}=\mathcal{M}^0_{xy}\frac{1}{2N}\;\mathrm{Re}\;\big[\omega^2d^*_{k\omega}\big(\Psi_{xk}^*(\omega)\Psi_{yl}(\omega)
-\Psi_{yk}^*(\omega)\Psi_{xl}(\omega)\big)d_{l\omega}\big]=0.
\end{equation}
Collecting terms we find that the matrix $\breve{\du{D}}^c(\omega)$ in Eq. (5.23) is given by
\begin{eqnarray}
\label{5.29}\breve{\du{D}}^c(\omega)&=&\du{D}^c+(\du{H}_c-a^{-2}\du{c}_0\cdot{\vc{\zeta}^0}')\cdot\du{u}_\beta\vc{\Psi}^*_c(\omega)\vc{\Psi}_\beta(\omega)
+a^{-1}\du{X}\cdot{\vc{\zeta}^0}'\nonumber\\
&-&(\du{G}_{c\beta}+a^{-1}\du{X}\cdot{\du{f}^0}'_\beta)\vc{\Psi}_\beta(\omega)+(\du{H}_c-a^{-2}\du{c}_0\cdot{\vc{\zeta}^0}'-\du{G}_{cc}-a^{-1}\du{X}\cdot{\du{f}^0}'_c)\vc{\Psi}_c(\omega).
\end{eqnarray}
The second order mean swimming velocities $\overline{U^{(2)}_x},\overline{U^{(2)}_y},\overline{\Omega^{(2)}}$ are calculated from the solution of Eqs. (4.6), (4.11) and (4.13). The expressions are complicated, but in our application they simplify because many of the coefficients vanish by symmetry.  We remark that the expressions in Eqs. (5.21) and (5.29) vanish in the absence of frictional hydrodynamic interactions. In that case the $F$-array in Eq. (5.16) is identically zero and the matrix ${\vc{\zeta}^0}'$ vanishes.

Finally we note that in Eqs. (5.14) and (5.23) the matrices $i\omega\breve{\du{D}}^\alpha(\omega)$ and $i\omega\breve{\du{D}}^c(\omega)$ can be reduced to their hermitian part. Also we can use projectors to take account of the fact that in swimming the displacement vector $\du{d}_\omega$ must be orthogonal to $\du{u}_x$, $\du{u}_y$ and $\hat{\du{c}}_0$.

\section{\label{VI}Triangular assembly}

As an application of the equations derived in the preceding sections we consider swimming of a triangular assembly of spheres of radius $a$ and mass density $\rho$. For simplicity we consider neutrally buoyant spheres. As basic equilibrium configuration $\du{c}_0$ we consider the equilateral triangle given by
\begin{equation}
\label{6.1}\du{c}_0=(-\frac{d}{2\sqrt{3}},-\frac{d}{2},\frac{d}{\sqrt{3}},0,-\frac{d}{2\sqrt{3}},\frac{d}{2}).
\end{equation}
The center of this triangle is at the origin and the three sides have length $d$. We have chosen the orientation such that with suitable displacements $\du{d}(t)$ the swimming is in the $x$ direction. The basic vectors $\du{u}_\alpha$ and $\hat{\du{c}}_0$ are
\begin{eqnarray}
\label{6.2}\du{u}_x&=&(1,0,1,0,1,0),\qquad
\du{u}_y=(0,1,0,1,0,1),\nonumber\\
\hat{\du{c}}_0&=&\frac{d}{a}(\frac{1}{2},-\frac{1}{2\sqrt{3}},0,\frac{1}{\sqrt{3}},-\frac{1}{2},-\frac{1}{2\sqrt{3}}).
\end{eqnarray}

We take the potential energy to be given by the expression
\begin{equation}
\label{6.3}V_{int}(\du{R})=\frac{k}{d^2}\big[(\vc{r}_{12}\cdot\vc{r}_{12}-d^2)^2+(\vc{r}_{23}\cdot\vc{r}_{23}-d^2)^2+(\vc{r}_{31}\cdot\vc{r}_{31}-d^2)^2\big],
\end{equation}
with elastic constant $k$ and relative distance vectors
\begin{equation}
\label{6.4}\vc{r}_{12}=\vc{R}_2-\vc{R}_1,\qquad\vc{r}_{23}=\vc{R}_3-\vc{R}_2,\qquad\vc{r}_{31}=\vc{R}_1-\vc{R}_3.
\end{equation}
The potential energy is positive definite, translation-invariant, isotropic, and it vanishes at configuration $\du{c}_0$.

We assume that the mobility matrix $\vc{\mu}(\du{R})$ is given by Oseen hydrodynamic interactions \cite{13}. Hence the friction matrix $\vc{\zeta}(\du{R})$ is evaluated to first order in the ratio $a/d$. We assume that the inverse mass matrix $\du{w}(\du{R})$ is evaluated in dipole approximation \cite{10}. Hence the mass matrix $\du{m}(\du{R})$ is evaluated to first order in the ratio $a^3/d^3$. The kinetic energy $\mathcal{K}(\du{R},\du{p})$ is positive definite, translation-invariant, and isotropic. The friction matrix and the mass matrix depend only on the relative distance vectors given by Eq. (6.4).

We introduce projection operators $\du{P}_{op}$ and $\du{Q}$ defined as
\begin{equation}
\label{6.5}\du{P}_{op}=\frac{1}{3}\du{u}_x\du{u}_x+\frac{1}{3}\du{u}_y\du{u}_y+\frac{\hat{\du{c}}_0\hat{\du{c}}_0}{\hat{\du{c}}_0\cdot\hat{\du{c}}_0},\qquad \du{Q}=\du{I}-\du{P}_{op}.
\end{equation}
The displacement vector $\du{d}_\omega$ must satisfy
\begin{equation}
\label{6.6}\du{d}_\omega=\du{Q}\cdot\du{d}_\omega,
\end{equation}
to exclude rigid body motion. The projected vector has only three independent components. This allows reduction of the matrices $\breve{\du{D}}^\alpha(\omega)$ and $\breve{\du{D}}^c(\omega)$ in Eqs. (5.21) and (5.29) to a three-dimensional representation.

In order to construct the reduced matrices we expand the displacement vector $\du{d}_\omega$ in terms of a convenient set of basis vectors. We use the orthonormal set of eigenvectors of the elasticity matrix corresponding to the interaction energy $V_{int}$. To second order in deviations from equilibrium we find from Eq. (6.3)
\begin{equation}
\label{6.7}V_{int2}=\frac{1}{2}\;(\du{R}-\du{R}_0)\cdot\du{H}\cdot(\du{R}-\du{R}_0),
\end{equation}
with $6\times 6$ elasticity matrix $\du{H}$ given explicitly by
 \begin{equation}
\label{6.8}\du{H}=k\left(\begin{array}{cccccc}6&2\sqrt{3}&-6&-2\sqrt{3}&0&0\\2\sqrt{3}&10&-2\sqrt{3}&-2&0&-8\\-6&-2\sqrt{3}&12&0&-6&2\sqrt{3}\\-2\sqrt{3}&-2&0&4&2\sqrt{3}&-2\\
0&0&-6&2\sqrt{3}&6&-2\sqrt{3}\\0&-8&2\sqrt{3}&-2&-2\sqrt{3}&10
\end{array}\right).
\end{equation}
This has the orthonormal set of eigenvectors
 \begin{eqnarray}
\label{6.9}\du{e}_1&=&(1,0,1,0,1,0)/\sqrt{3}=\du{u}_x/\sqrt{3},\nonumber\\
\du{e}_2&=&(0,1,0,1,0,1)/\sqrt{3}=\du{u}_y/\sqrt{3},\nonumber\\
\du{e}_3&=&\big(\frac{1}{2},\frac{-1}{2\sqrt{3}},0,\frac{1}{\sqrt{3}},\frac{-1}{2},\frac{-1}{2\sqrt{3}}\big)=\frac{a}{d}\;\hat{\du{c}}_0,\nonumber\\
\du{e}_4&=&\big(\frac{1}{2},\frac{-1}{2\sqrt{3}},\frac{-1}{2},\frac{-1}{2\sqrt{3}},0,\frac{1}{\sqrt{3}}\big),\nonumber\\
\du{e}_5&=&\big(\frac{-1}{2\sqrt{3}},\frac{-1}{2},\frac{-1}{2\sqrt{3}},\frac{1}{2},\frac{1}{\sqrt{3}},0\big),\nonumber\\
\du{e}_6&=&\big(\frac{-1}{2\sqrt{3}},\frac{-1}{2},\frac{1}{\sqrt{3}},0,\frac{-1}{2\sqrt{3}},\frac{1}{2}\big)=\frac{1}{d}\;\du{c}_0,
\end{eqnarray}
with eigenvalues
\begin{equation}
\label{6.10}\lambda_1=\lambda_2=\lambda_3=0,\qquad\lambda_4=\lambda_5=12k,\qquad\lambda_6=24k.
\end{equation}
The first three eigenvectors and eigenvalues correspond to free translation and rotation.

The basis vectors $\du{e}_4,\du{e}_5,\du{e}_6$ span the reduced vector space. The corresponding displacement vector
\begin{equation}
\label{6.11}\du{d}_\omega=b_4\du{e}_4+b_5\du{e}_5+b_6\du{e}_6
\end{equation}
is characterized by three complex coefficients $b_4,b_5,b_6$ in complex notation.

The matrices $\du{D}^\alpha$ and $\du{D}^c$ do not depend on frequency. We can express the matrix $\du{D}^x$ at $\du{R}_0$ as
 \begin{eqnarray}
\label{6.12}\du{D}^x=\frac{9\pi\eta a^2}{16d^2}\bigg[4 \sqrt{3}\;\du{e}_3\du{e}_4-12\du{e}_3\du{e}_5+6\du{e}_4\du{e}_1+2\sqrt{3}\;\du{e}_4\du{e}_2
-6\sqrt{3}\;\du{e}_4\du{e}_3+3\sqrt{3}\;\du{e}_4\du{e}_4+9\du{e}_4\du{e}_5\nonumber\\
+2\sqrt{3}\;\du{e}_5\du{e}_1-6\du{e}_5\du{e}_2+18\du{e}_5\du{e}_3+9\du{e}_5\du{e}_4-3\sqrt{3}\;\du{e}_5\du{e}_5
+24\sqrt{3}\;\du{e}_6\du{e}_1-6\du{e}_6\du{e}_4-2\sqrt{3}\;\du{e}_6\du{e}_5\bigg].\nonumber\\
\end{eqnarray}
Hence the reduced $3\times 3$ matrix $\du{D}^x_3$
is given by
 \begin{equation}
\label{6.13}\du{D}^x_3=\frac{9\pi\eta a^2}{16d^2}\left(\begin{array}{ccc}3\sqrt{3}&9&0\\9&-3\sqrt{3}&0\\-6&-2\sqrt{3}&0
\end{array}\right).
\end{equation}
The matrix $\du{D}^y$ at $\du{R}_0$ can be expressed as
 \begin{eqnarray}
\label{6.14}\du{D}^y=\frac{9\pi\eta a^2}{16d^2}\bigg[12\du{e}_3\du{e}_4+4\sqrt{3}\;\du{e}_3\du{e}_5+2\sqrt{3}\;\du{e}_4\du{e}_1-6\du{e}_4\du{e}_2
-18\du{e}_4\du{e}_3+9\du{e}_4\du{e}_4-3\sqrt{3}\;\du{e}_4\du{e}_5\nonumber\\
-6\du{e}_5\du{e}_1-2\sqrt{3}\;\du{e}_5\du{e}_2-6\sqrt{3}\;\du{e}_5\du{e}_3-3\sqrt{3}\;\du{e}_5\du{e}_4-9\du{e}_5\du{e}_5
+24\sqrt{3}\;\du{e}_6\du{e}_2+2\sqrt{3}\;\du{e}_6\du{e}_4-6\du{e}_6\du{e}_5\bigg].\nonumber\\
\end{eqnarray}
Hence the reduced $3\times 3$ matrix $\du{D}^y_3$
is given by
 \begin{equation}
\label{6.15}\du{D}^y_3=\frac{9\pi\eta a^2}{16d^2}\left(\begin{array}{ccc}9&-3\sqrt{3}&0\\-3\sqrt{3}&-9&0\\2\sqrt{3}&-6&0
\end{array}\right).
\end{equation}
The matrix $\du{D}^c$ at $\du{R}_0$ can be expressed as
 \begin{equation}
\label{6.16}\du{D}^c=\frac{9\pi\eta a}{8d}\bigg[8\du{e}_3\du{e}_6-3\du{e}_4\du{e}_1-3\sqrt{3}\;\du{e}_4\du{e}_2
+7\du{e}_4\du{e}_5+3\sqrt{3}\;\du{e}_5\du{e}_1-3\du{e}_5\du{e}_2-7\du{e}_5\du{e}_4-2\du{e}_6\du{e}_3\bigg].
\end{equation}
Hence the reduced $3\times 3$ matrix $\du{D}^c_3$
is given by
 \begin{equation}
\label{6.17}\du{D}^c_3=\frac{9\pi\eta a}{8d}\left(\begin{array}{ccc}0&7&0\\-7&0&0\\0&0&0
\end{array}\right).
\end{equation}

The coefficients in Eq. (5.21) take simple values. We find at $\du{R}_0$
 \begin{eqnarray}
\label{6.18}\du{H}_\alpha\cdot\du{u}_\beta&=&0,\qquad \du{G}_{xx}=\frac{27\pi\eta a^2}{8d^2}\big(\sqrt{3}\;\du{e}_4+\du{e}_5+12\du{e}_6\big),\nonumber\\
\du{G}_{xy}&=&\frac{27\pi\eta a^2}{8d^2}\big(\du{e}_4-\sqrt{3}\;\du{e}_5\big),\qquad \du{G}_{yy}=\frac{27\pi\eta a^2}{8d^2}\big(-\sqrt{3}\;\du{e}_4-\du{e}_5+12\du{e}_6\big)\nonumber\\
\du{H}_x-\du{G}_{xc}&=&\frac{45\pi\eta a}{8d}\big(\sqrt{3}\;\du{e}_4-3\du{e}_5\big),\qquad \du{H}_y-\du{G}_{yc}=\frac{45\pi\eta a}{8d}\big(3\du{e}_4+\sqrt{3}\;\du{e}_5\big).
\end{eqnarray}
Similarly we find for the coefficients in Eq. (5.29) at $\du{R}_0$
 \begin{eqnarray}
\label{6.19} \big(\du{H}_c-a^{-2}\du{c}_0\cdot{\vc{\zeta}^0}'\big)\cdot\du{u}_\beta&=&0,\nonumber\\
\qquad \du{G}_{cx}+a^{-1}\du{X}\cdot{\du{f}^0}'_x&=&\frac{9\pi\eta a}{2d}\big(3\sqrt{3}\;\du{e}_2-\sqrt{3}\;\du{e}_4+3\du{e}_5\big),\nonumber\\
\du{G}_{cy}+a^{-1}\du{X}\cdot{\du{f}^0}'_y&=&\frac{9\pi\eta a}{2d}\big(-3\sqrt{3}\;\du{e}_1-3\du{e}_4-\sqrt{3}\;\du{e}_5\big),\nonumber\\\du{H}_c-a^{-2}\du{c}_0\cdot{\vc{\zeta}^0}'-\du{G}_{cc}&=&0,\qquad
a^{-1}\du{X}\cdot{\du{f}^0}'_c=\frac{9\pi\eta}{4}\du{e}_6.
\end{eqnarray}
We also have
\begin{eqnarray}
\label{6.20}a^{-1}\du{X}\cdot{\vc{\zeta}^0}'=\frac{9\pi\eta a}{8d}\bigg[-12\du{e}_1\du{e}_2-\du{e}_1\du{e}_4
+\sqrt{3}\;\du{e}_1\du{e}_5+12\du{e}_2\du{e}_1-\sqrt{3}\;\du{e}_2\du{e}_4-\du{e}_2\du{e}_5\nonumber\\
-10\du{e}_3\du{e}_6-\du{e}_4\du{e}_1-\sqrt{3}\;\du{e}_4\du{e}_2-6\du{e}_4\du{e}_5+\sqrt{3}\;\du{e}_5\du{e}_1-\du{e}_5\du{e}_2+6\du{e}_5\du{e}_4+2\du{e}_6\du{e}_3\bigg].
\end{eqnarray}

The elements of the admittance matrix take simple values. For the mass elements defined in Eq. (5.4) we find
\begin{eqnarray}
\label{6.21}\mathcal{M}^0_{xx}&=&\mathcal{M}^0_{yy}=6\pi a^3\rho\bigg[1-\frac{a^3}{2d^3}\bigg],\qquad\mathcal{M}^0_{xy}=\mathcal{M}^0_{yx}=0,\nonumber\\
\mathcal{M}^0_{xc}&=&\mathcal{M}^0_{cx}=\mathcal{M}^0_{yc}=\mathcal{M}^0_{cy}=0,\qquad\mathcal{M}^0_{cc}=2\pi ad^2\rho\bigg[1-\frac{5a^3}{4d^3}\bigg].
\end{eqnarray}
For the friction elements we find
\begin{eqnarray}
\label{6.22}Z^0_{xx}&=&Z^0_{yy}=18\pi\eta a\bigg[1-\frac{9a}{4d}\bigg],\qquad Z^0_{xy}=Z^0_{yx}=0,\nonumber\\
Z^0_{xc}&=&Z^0_{cx}=Z^0_{yc}=Z^0_{cy}=0,\qquad Z^0_{cc}=6\pi\eta\frac{d^2}{a}\bigg[1+\frac{3a}{8d}\bigg].
\end{eqnarray}

The impedance vectors defined in Eq. (5.8) are easily evaluated. Hence we find for the elements defined in Eq. (5.10)
 \begin{eqnarray}
\label{6.23}\vc{\Psi}_x(\omega)&=&\frac{1}{\sqrt{3}}\;\du{e}_1+\frac{3}{4}\;S\du{e}_4+\frac{\sqrt{3}}{4}\;S\du{e}_5,\nonumber\\
\vc{\Psi}_y(\omega)&=&\frac{1}{\sqrt{3}}\;\du{e}_2-\frac{\sqrt{3}}{4}\;S\du{e}_4+\frac{3}{4}\;S\du{e}_5,\nonumber\\
\vc{\Psi}_c(\omega)&=&\frac{a}{d}\;\du{e}_3,
\end{eqnarray}
with coefficient
 \begin{equation}
\label{6.24}S=\frac{3ad^2-4ia^3s^2}{12d^3-8is^2d^3-27ad^2+4ia^3s^2},
\end{equation}
with scale number $s$ defined by the dimensionless ratio \cite{4}
 \begin{equation}
\label{6.25}s^2=\frac{a^2\omega\rho}{2\eta}.
\end{equation}
The value $s=0$ corresponds to the Stokes limit, and large $s$ corresponds to the inertia-dominated regime.
Collecting terms we find for the reduced matrix $\breve{\du{D}}^x_3$
 \begin{equation}
\label{6.26}\breve{\du{D}}^x_3=\du{D}^x_3-\frac{27\pi\eta a^2}{16d^2}\;S\;\left(\begin{array}{ccc}-\sqrt{3}&3&0\\3&-\sqrt{3}&0\\18&6\sqrt{3}&0
\end{array}\right).
\end{equation}
Similarly we find for the reduced matrix $\breve{\du{D}}^y_3$
 \begin{equation}
\label{6.27}\breve{\du{D}}^y_3=\du{D}^y_3+\frac{27\pi\eta a^2}{16d^2}\;S\;\left(\begin{array}{ccc}-3&\sqrt{3}&0\\\sqrt{3}&3&0\\-6\sqrt{3}&18&0
\end{array}\right).
\end{equation}
For the reduced matrix $\breve{\du{D}}^c_3$ we find
 \begin{equation}
\label{6.28}\breve{\du{D}}^c_3=\du{D}^c_3+\frac{27\pi\eta a}{4d}\left(\begin{array}{ccc}0&-1+2S&0\\1-2S&0&0\\0&0&0
\end{array}\right).
\end{equation}
The matrices can be used in Eqs. (5.14) and (5.23) to evaluate the mean translational and rotational impetus for complex displacement vector of the form (6.11).

\section{\label{VII}Swimming triangular assembly}

In this section we evaluate the mean translational and rotational swimming velocity of a triangular assembly of spheres. Besides the mean translational and rotational impetus we need the generalized ellipticity defined in Eq. (4.11). We evaluate this for complex displacement vectors of the form (6.11). We note that
 \begin{equation}
\label{7.1}\du{e}_4\cdot\du{X}\cdot\du{e}_5=-\du{e}_5\cdot\du{X}\cdot\du{e}_4=-1.
\end{equation}
All other cross products between basis vectors $\du{e}_4,\du{e}_5,\du{e}_6$ vanish. Hence we find from Eq. (4.11)
\begin{equation}
\label{7.2}\overline{\mathcal{E}}=B_{45}/a^2,
\end{equation}
with the abbreviation
\begin{equation}
\label{7.3}B_{45}=\mathrm{Im}\;[b_4b^*_5].
\end{equation}
It follows from Eq. (6.22) that Eqs. (4.6) and (4.13) decouple. We consider first Eq. (4.13) and combine this with Eq. (4.11). The mean rotational impetus is found from Eqs. (5.23), (6.17) and (6.28). By use of Eq. (4.13) this yields
\begin{equation}
\label{7.4}\alpha Z^0_{cc}a\overline{\Omega^{(2)}}=\frac{9\pi\eta a}{8d}\;\omega(1+2S')B_{45},
\end{equation}
where $S'$ is the real part of $S$ in Eq. (6.24). From Eqs. (4.11) and (7.2) we have
\begin{equation}
\label{7.5}(1-\alpha)Z^0_{cc}a^2\overline{\Omega^{(2)}}=\zeta_0\omega B_{45}.
\end{equation}
From the ratio of Eqs. (7.4) and (7.5) we obtain
\begin{equation}
\label{7.6}\alpha=\frac{\beta}{1+\beta},\qquad \beta=\frac{3a}{16d}\;[1+2S'].
\end{equation}
 To lowest order in the ratio $a/d$ this becomes $\alpha=3a/(3a+16d)$. We write
 \begin{equation}
\label{7.7}\overline{\Omega^{(2)}}=\overline{\Omega^{(2)}}_0+{\overline{\Omega^{(2)}}}\;',\qquad\overline{\Omega^{(2)}}_0=(1-\alpha)\overline{\Omega^{(2)}},\qquad{\overline{\Omega^{(2)}}}\;'
=\alpha\overline{\Omega^{(2)}},
\end{equation}
with the first term given by Eq. (7.5) and the second by Eq. (7.4), and identify ${\overline{\Omega^{(2)}}}\;'$ as the mean rotational swimming velocity. We find
 \begin{equation}
\label{7.8}\overline{\Omega^{(2)}}_0=\frac{B_{45}}{d(3a+8d)}\;\omega,
\qquad\overline{\Omega^{(2)}}\;'=\beta\frac{B_{45}}{d(3a+8d)}\;\omega.
\end{equation}
In Fig. 1 we plot the factor $1+2S'$ as a function of square scale number $s^2$ for $d=3a$.

The mean translational impetus is found from Eq. (5.14). By use of Eq. (4.6) this yields for the $x$ and $y$ components of the second order mean swimming velocity
\begin{equation}
\label{7.9}\overline{U_\alpha^{(2)}}=\frac{1}{2Z^0_{\alpha\alpha}}\;\mathrm{Re}\;[i\omega\du{b}_\omega^*\cdot\breve{\du{D}}^\alpha_3\cdot\du{b}_\omega],\qquad (\alpha=x,y).
\end{equation}
with complex 3-dimensional vector $\du{b}_\omega$ with components $b_4,b_5,b_6$ as in Eq. (6.11). Explicitly the $x$ component is given by
\begin{eqnarray}
\label{7.10}\overline{U^{(2)}_x}=\frac{a\omega}{16d(4d-9a)}\bigg[6\;\mathrm{Im}[b_4b_6^*]+2\sqrt{3}\;\mathrm{Im}[b_5b_6^*]\nonumber\\
-\big(3\sqrt{3}|b_4|^2-9b_4b_5^*-9b_4^*b_5+3\sqrt{3}|b_5|^2\big)S''\nonumber\\
+\mathrm{Im}\big[(54b_4b_6^*+18\sqrt{3}b_5b_6^*)(S'+iS'')\big]\bigg],
\end{eqnarray}
and the $y$ component is given by
\begin{eqnarray}
\label{7.11}\overline{U^{(2)}_y}=\frac{a\omega}{16d(4d-9a)}\bigg[-2\sqrt{3}\;\mathrm{Im}[b_4b_6^*]+6\;\mathrm{Im}[b_5b_6^*]\nonumber\\
-(-9\big(|b_4|^2+3\sqrt{3}b_4b_5^*+3\sqrt{3}b_4^*b_5+9|b_5|^2\big)S''\nonumber\\
+\mathrm{Im}\big[(18\sqrt{3}b_4b_6^*-54b_5b_6^*)(S'+iS'')\big]\bigg].
\end{eqnarray}
It is seen from Eq. (6.24) that in the Stokes limit and in the inertia-dominated regime $S''=0$, so that in these limits the translational swimming velocity is a consequence of coupling between the modes labeled 4 and 6, or 5 and 6.

If we choose the ratio $b_4/b_5$ to be real, then $B_{45}$, defined in Eq. (7.3), vanishes. In this case both $\overline{U^{(2)}_x}$ and $\overline{U^{(2)}_y}$ in general are non-vanishing and the swimmer performs a circular motion with vanishing $\overline{\Omega^{(2)}}_0$ and $\overline{\Omega^{(2)}}\;'$. For suitable choice of the real ratio $b_4/b_5$ the swimmer on average moves in the $x$ direction without rotation. In this situation the mean swimming velocity $\overline{U^{(2)}_x}$ for given power can be optimized by suitable choice of the complex ratio $b_4/b_6$.

\section{\label{VIII}Power and optimization}

Finally we consider the question how to optimize the stroke such that the mean translational swimming velocity is maximal for given power. From Eq. (3.11) the power is identified as the mean rate of dissipation during a period. As before we consider small amplitude swimming and evaluate the power to second order in the amplitude. The optimization leads to a generalized eigenvalue problem involving two hermitian matrices.

The time-dependent rate of dissipation is given by Eq. (3.9). The second order mean rate of dissipation can be expressed in terms of periodic displacements with Fourier amplitude $\du{d}_\omega$ as
\begin{equation}
\label{8.1}\overline{\mathcal{D}^{(2)}}=\frac{1}{2}\;\omega^2\mathrm{Re}\;[\du{d}_\omega^*\cdot\breve{\vc{\zeta}}(\omega)\big|_{\du{R}_0}\cdot\du{d}_\omega],
\end{equation}
with modified friction matrix \cite{10},\cite{18}
\begin{equation}
\label{8.2}\breve{\vc{\zeta}}(\omega)=\vc{\zeta}-Y_{\rho\sigma}(\omega)\du{f}_\rho(\omega)\du{f}_\sigma(\omega).
\end{equation}
It suffices to consider the hermitian part of the matrix. We define the hermitian matrix $\du{A}(\omega)$ as
\begin{equation}
\label{8.3}\du{A}(\omega)=\frac{1}{2\eta a}\big[\breve{\vc{\zeta}}(\omega)+\breve{\vc{\zeta}}(\omega)^\dagger\big].
\end{equation}
The elements of $\du{A}(\omega)$ are dimensionless. For the triangular swimmer in the representation given by Eq. (6.9) the elements of the first three rows and columns of the six-dimensional matrix $\du{A}(\omega)$ vanish. Therefore we can restrict attention to the $3\times 3$ matrix $\du{A}_3(\omega)$ obtained by deleting the first three rows and columns. The matrix $\du{A}_3(\omega)$ is diagonal. In the Stokes limit $s=0$ the diagonal elements are
\begin{equation}
\label{8.4}A_{344}(0)=A_{355}(0)=3\pi\frac{64d^2-72ad-171a^2}{8d(4d-9a)},\qquad A_{366}(0)=3\pi\bigg(2+\frac{15a}{4d}\bigg).
\end{equation}

From Eq. (7.9) we define correspondingly
\begin{equation}
\label{8.5}\du{B}_3^\alpha(\omega)=\frac{ia}{2Z^0_{\alpha\alpha}}\;\big(\breve{\du{D}}^\alpha_3-\breve{\du{D}}^\alpha_3\;^\dagger\big),\qquad (\alpha=x,y).
\end{equation}
In the Stokes limit $s=0$ the explicit expressions are
 \begin{equation}
\label{8.6}\du{B}^x_3(0)=\frac{ia^2}{2(4d-9a)^2}\left(\begin{array}{ccc}0&0&3\\0&0&\sqrt{3}\\-3&-\sqrt{3}&0
\end{array}\right),\qquad\du{B}^y_3(0)=\frac{ia^2}{2(4d-9a)^2}\left(\begin{array}{ccc}0&0&-\sqrt{3}\\0&0&3\\\sqrt{3}&-3&0
\end{array}\right).
\end{equation}

The optimal stroke for swimming in the $x$ direction for given power is given by the eigenvector of the generalized eigenvalue problem
\begin{equation}
\label{8.7}\du{B}_3^x(\omega)\cdot\du{b}=\lambda\du{A}_3(\omega)\cdot\du{b},
\end{equation}
with maximum eigenvalue $\lambda_{max}$. The maximum eigenvalue equals the efficiency $E_T=\eta\omega a^2|\overline{U^{(2)}_x}|/\overline{\mathcal{D}^{(2)}}$ of the stroke $\du{b}$.
In the Stokes limit $s=0$ we find from the above expressions
\begin{equation}
\label{8.8}\lambda_{max}(0)=\frac{1}{3\pi}\frac{4\sqrt{6}\;a^2d}{[(4d-9a)^3(8d+15a)(64d^2-72ad-171a^2)]^{1/2}},
\end{equation}
with corresponding eigenvector
\begin{equation}
\label{8.9}\du{b}=(1,\;1/\sqrt{3},\;-i\;\frac{64d^2-72ad-171a^2}{18a^2}\;\lambda_{max}(0)),
\end{equation}
normalized such that the first component equals unity. It is remarkable that the second component is independent of the ratio $d/a$. For general values of the scale number $s$ we also obtain analytic expressions. In Fig. 2 we show the maximum eigenvalue $\lambda_{max}(s)$ for $d=3a$ as a function of $s$.

It is evident from Eq. (8.9) that in the Stokes limit the ratio $b_4/b_5$ is real, so that $B_{45}$ in Eq. (7.3) vanishes. Correspondingly the mean ellipticity in Eq. (7.2) and the mean rotational swimming velocity $\overline{\Omega^{(2)}}\;'$ in Eq. (7.8) vanish. We find that for the optimal stroke $\du{b}^*\cdot\du{B}^y_3(0)\cdot\du{b}=0$, so that the mean translational swimming velocity is in the $x$ direction. These statements hold also for general values of $s$.

It is worthwhile to compare the efficiency of the triangular swimmer in the Stokes limit with that of the three-sphere linear chain. In Fig. 3 we compare the maximum eigenvalue $\lambda_{max}$ for $s=0$ as a function of the ratio $d/a$ with that of the three-sphere chain \cite{9}. For $d<3.930a$ the triangular swimmer is more efficient than the three-sphere chain, at $d=3a$ by a factor $2.784$. For the triangular swimmer $\lambda_{max}=0.00699$ at $d=3a$, whereas for the three-sphere chain $\lambda_{max}=0.00251$ at $d=3a$.

The time-dependent displacement vector $\du{d}(t)$ corresponding to the eigenvector with eigenvalue $\lambda_{max}$ has Cartesian components with the properties
\begin{equation}
\label{8.10}d_{1x}(t)=d_{3x}(t),\qquad d_{1y}(t)=-d_{3y}(t),\qquad d_{2x}(t)=-2d_{1x}(t),\qquad d_{2y}(t)=0,
\end{equation}
indicating a symmetric motion. The centers of sphere 1 and 3 run through an elliptical orbit in the $xy$ plane during a period. The center of sphere 2 moves back and forth along the $x$ axis, such that the center of mass remains fixed. In Fig. 4 we show the elliptical orbit of the center of sphere 1 for $d=3a,\;s=0,$ and arbitrarily chosen amplitude.

\section{\label{IX}Discussion}

The above analysis of swimming at small Reynolds number is instructive, since it shows how to go beyond the treatment of the usually studied collinear systems. The planar geometry provides an additional degree of freedom, corresponding to translational swimming in two directions, as well as to rotational swimming. This type of self-propulsion was studied by Vladimirov \cite{11} in the Stokes limit for three spheres connected by rods which change their length independently and periodically.

Our analysis of the translational and rotational swimming velocity is based on balance equations for momentum and angular momentum of the assembly. Since these do not contain the actuating forces explicitly we can use a kinematic approach in which the displacements from equilibrium positions are prescribed as periodic functions of time. We evaluated the mean swimming velocities and the mean power to second order in the amplitude of displacements. This allows optimization of the translational swimming velocity at given power by means of a generalized eigenvalue problem.

In principle the analysis can be extended to fully three-dimensional situations, though at the expense of additional complication. It seems more attractive to consider the limited extension to planar structures with displacements perpendicular to the plane, as well as in the plane. This would allow to study the effect of flapping motion.

At larger amplitude of stroke the motion of the spheres would cause vortex shedding \cite{6A}. This effect is absent from the resistive-reactive model studied here. It would be of interest to compare the results calculated for the model with those obtained by computer simulation.\\

$\mathrm{\bf{Acknowledgment}}$ I thank Professor V. A. Vladimirov for stimulating correspondence.

\newpage

\section*{Figure captions}

\subsection*{Fig. 1}
Plot of the factor $1+2S'$ in Eq. (7.8) as a function of square scale number $s^2$ for $d=3a$.

\subsection*{Fig. 2}
Plot of the ratio $\lambda_{max}(s)/\lambda_{max}(0)$ for $d=3a$ as a function of scale number $s$.

\subsection*{Fig. 3}
Plot of the maximum eigenvalue $\lambda_{max}$ for $s=0$ as a function of the ratio $d/a$ for the triangle (drawn curve) compared with that for the three-sphere linear chain (dashed curve).
\subsection*{Fig. 4}
Plot of the elliptical orbit of the center of sphere 1 in the $xy$ plane for $d=3a$ for the optimal stroke in the Stokes limit. The amplitude is chosen arbitrarily.

\newpage
\setlength{\unitlength}{1cm}
\begin{figure}
 \includegraphics{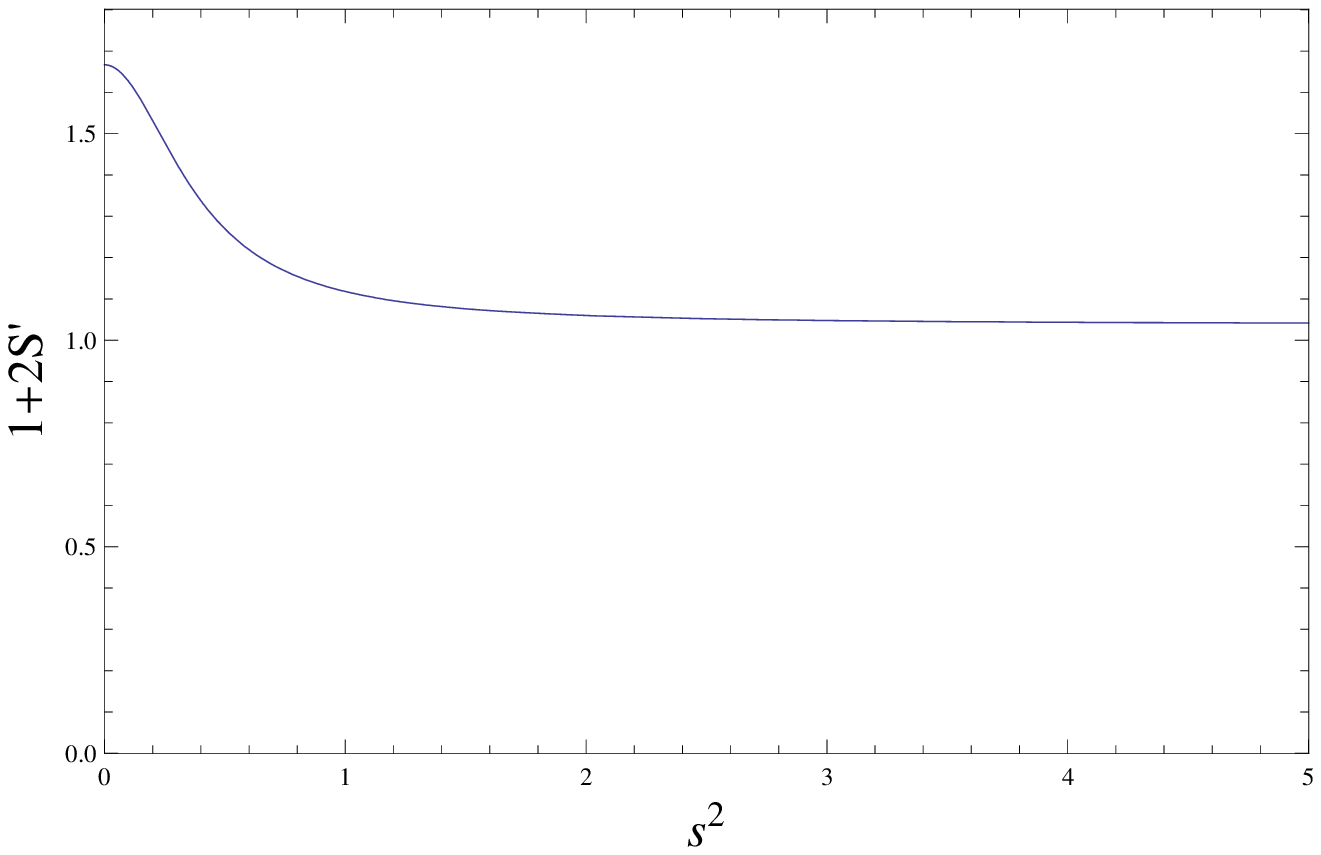}
   \put(-9.1,3.1){}
\put(-1.2,-.2){}
  \caption{}
\end{figure}
\newpage
\clearpage
\newpage
\setlength{\unitlength}{1cm}
\begin{figure}
 \includegraphics{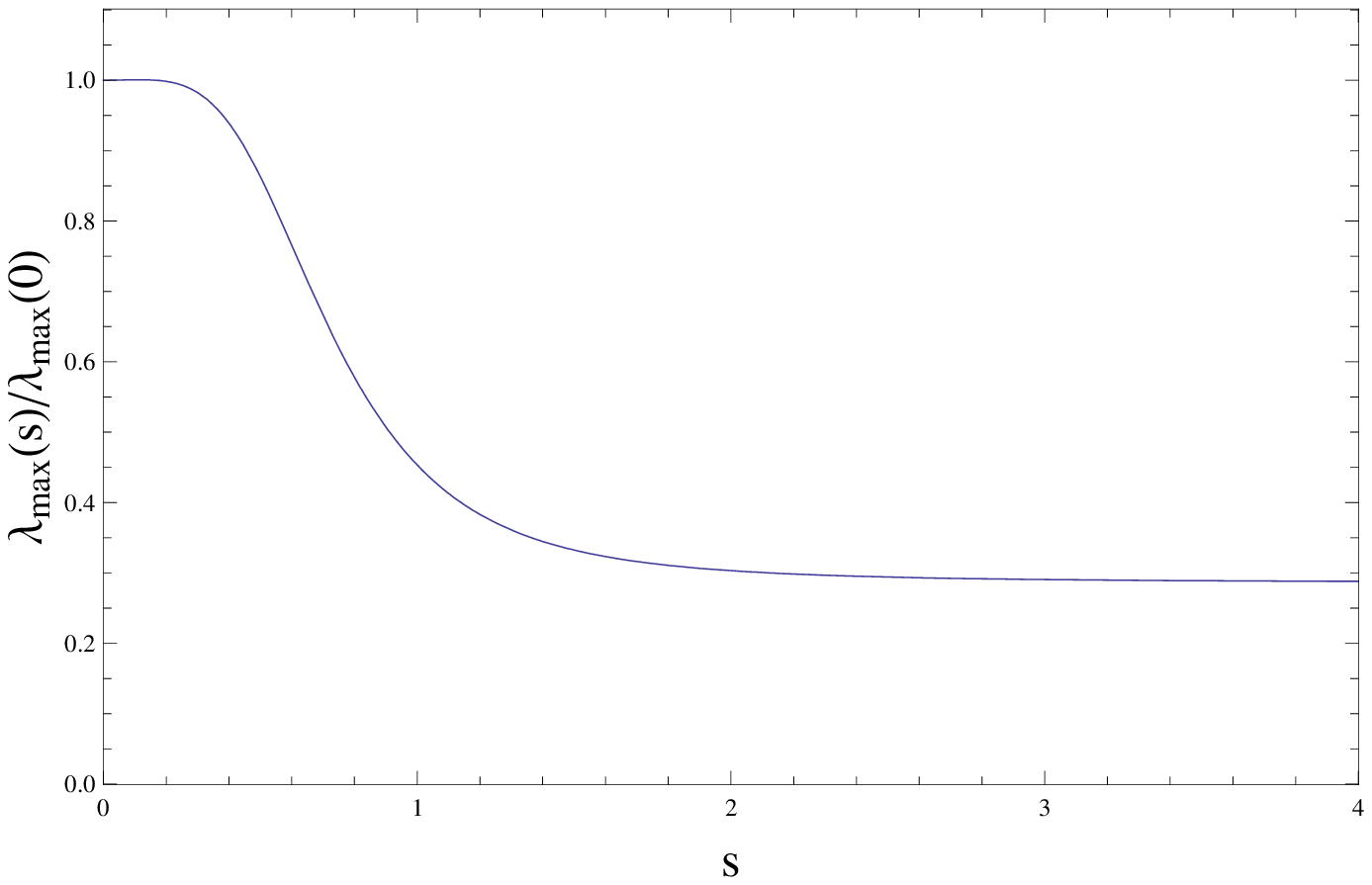}
   \put(-9.1,3.1){}
\put(-1.2,-.2){}
  \caption{}
\end{figure}
\newpage
\clearpage
\newpage
\setlength{\unitlength}{1cm}
\begin{figure}
 \includegraphics{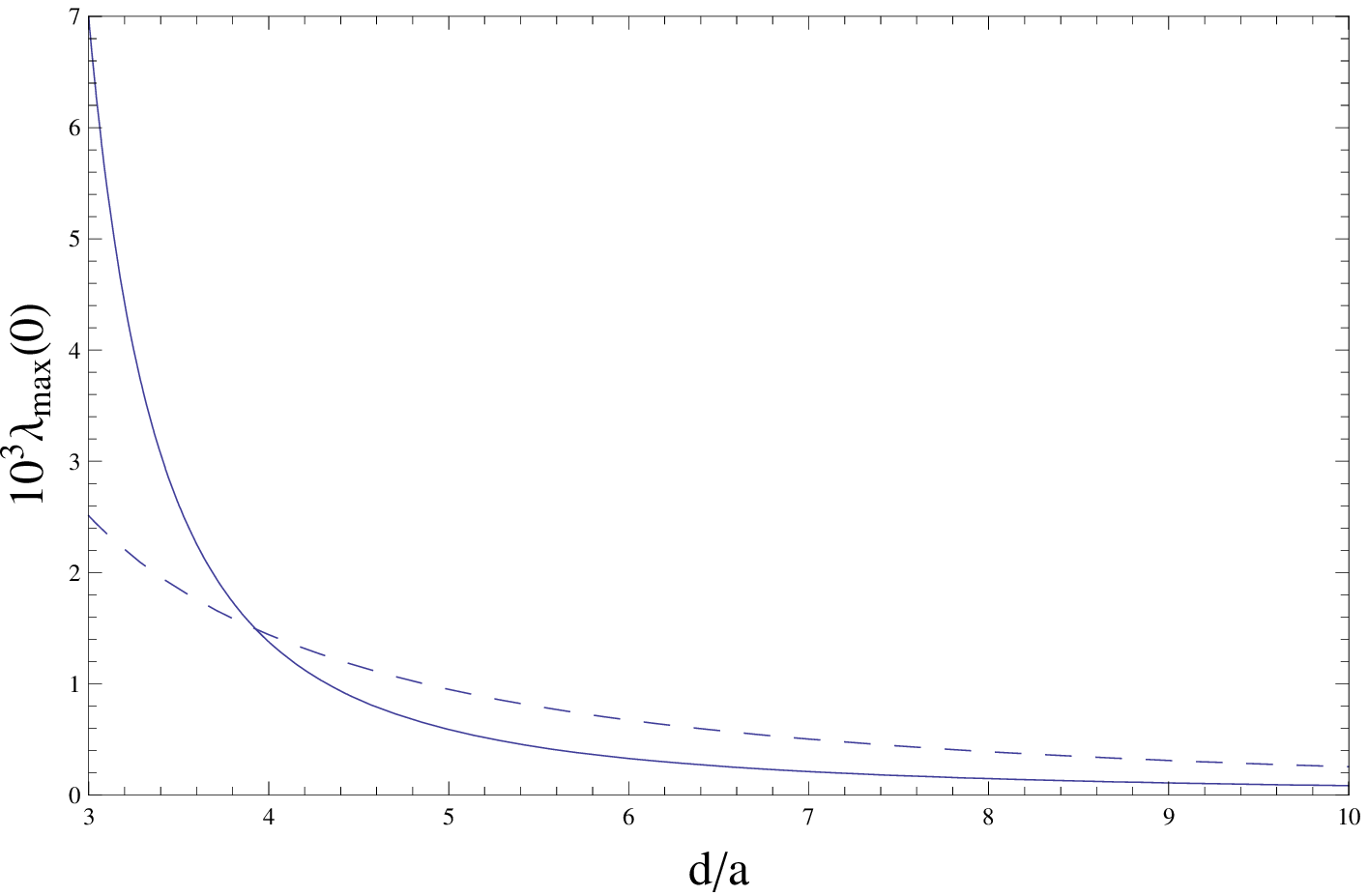}
   \put(-9.1,3.1){}
\put(-1.2,-.2){}
  \caption{}
\end{figure}
\newpage
\clearpage
\newpage
\setlength{\unitlength}{1cm}
\begin{figure}
 \includegraphics{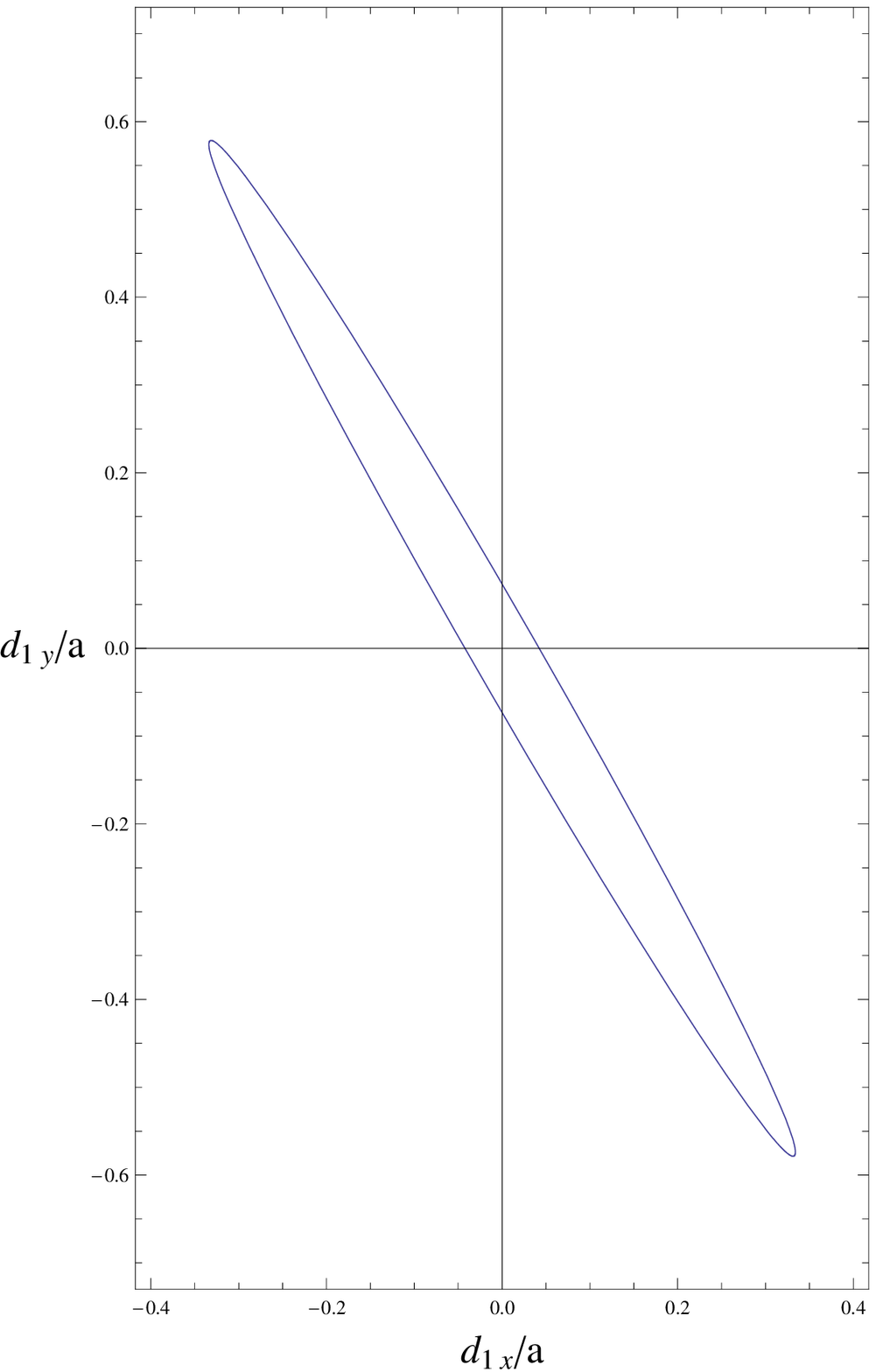}
   \put(-9.1,3.1){}
\put(-1.2,-.2){}
  \caption{}
\end{figure}
\newpage

\newpage


\begin{thebibliography}{99}

\bibitem{1}
E. M. Purcell, "Life at low Reynolds number", Amer. J. of Phys. \vol{45}, 3 (1977).

\bibitem{2}
E. Lauga and T. R. Powers, "The hydodynamics of swimming microorganisms", Rep. Prog. Phys. \vol{72}, 09660 (2009).

\bibitem{3}
A. S. Sangani and A. Prosperetti, "Numerical simulation of the motion of particles at large Reynolds numbers", in {\it Particulate two-phase flow} ed. M. C. Roco (Butterworth-Heinemann, Boston, 1993).

\bibitem{4}
B. U. Felderhof and R. B. Jones, "Swimming of a sphere in a viscous incompressible fluid with inertia", arXiv:1512.04667[physics.flu-dyn].

\bibitem{5}
T. Y. Wu, "On Theoretical Modeling of Aquatic and Aerial Animal Locomotion", Adv. Appl. Mech. \vol{38}, 291 (2001).

\bibitem{6}
T. Y. Wu, "Fish Swimming and Bird/Insect Flight", Annu. Rev. Fluid Mech. \vol{43}, 25 (2011).

\bibitem{6A}
W. Shyy, Y. Lian, J. Tang, D. Viieru, and H. Liu, {\it Aerodynamics of low Reynolds number flyers}, Cambridge University Press, Cambridge, 2011).

\bibitem{7}
C. Caspersen, P. A. Berthelsen, M. Eik, C. P\^akozdi, and P.-L. Kjendlie, "Added mass in human swimmers: Age and gender differences", J. Biomech. \vol{43}, 2369 (2010).

\bibitem{8}
R. Golestanian and A. J. Ajdari, "Analytic results for the three-sphere swimmer at low Reynolds number", Phys. Rev. E \vol{77}. 036308 (2008).

\bibitem{9}
B. U. Felderhof, "Swimming of an assembly of rigid spheres at low Reynolds number", Eur. Phys. J. E \vol{37}, 110 (2014).

\bibitem{10}
B. U. Felderhof, "Effect of inertia on laminar swimming and flying of an assembly of rigid spheres in an incompressible viscous fluid", Phys. Rev. E \vol{92}, 053011 (2015).

\bibitem{10A}
B. U. Felderhof, "Swimming at small Reynolds number of a linear assembly of spheres in an incompressible viscous fluid with inertia", arXiv:161019.01266[physics.flu-dyn].

\bibitem{11}
V. A. Vladimirov, "Theory of a triangular micro-robot", arXiv:1210.0747[physics.flu-dyn].

\bibitem{12}
B. Cichocki, B. U. Felderhof, K. Hinsen, E. Wajnryb, and J. Blawzdziewicz, "Friction and mobility of many spheres in Stokes flow", J. Chem. Phys. \vol{100}, 3780 (1994).

\bibitem{13}
H. Yamakawa, {\it Modern Theory of Polymer Solutions} (Harper and Row, New York, 1971).

\bibitem{14}
J. Rotne and S. Prager, "Variational treatment of hydrodynamic interaction in polymers", J. Chem. Phys. \vol{50}, 4831 (1969).

\bibitem{15}
H. Lamb, {\it Hydrodynamics} (Cambridge University Press, Cambridge, 1932).

\bibitem{16}
J. Lighthill, {\it An Informal Introduction to Theoretical Fluid Mechanics} (Clarendon Press, Oxford, 1986).

\bibitem{17}
G. J. van Ingen Schenau and P. R. Cavanagh, "Power equations in endurance sports", J. Biomech. \vol{23}, 865 (1990).

\bibitem{18}
B. U. Felderhof, "Swimming of a linear chain with a cargo in an incompressible viscous fluid with inertia", arXiv:1607.08048[physics.flu-dyn].

\end{thebibliography}
\end{document}